\documentclass[prl,twocolumn,showpacs]{revtex4}

\usepackage{graphicx}
\usepackage{amsmath}

\begin{document}

\title{Phase Evolution in Spatial Dark States}

\author{S.~McEndoo$^{1}$, S.~Croke$^{2}$, J.~Brophy$^{1}$ and
  Th.~Busch$^{1}$}

\affiliation{
  $^{1}$Physics Department, University College Cork, Cork, Ireland\\
  $^{2}$Perimeter Institute for Theoretical Physics, Waterloo,
  Ontario, N2L 2Y5, Canada}

\date{\today}

\begin{abstract}
  Adiabatic techniques using multi-level systems have recently been
  generalised from the optical case to settings in atom
  optics, solid state and even classical electrodynamics. The most
  well known example of these is the so called STIRAP process, which
  allows transfer of a particle between different states with large
  fidelity. Here we generalise and examine this process for an atomic
  centre-of-mass state with a non-trivial phase distribution and show
  that even though dark state dynamics can be achieved for the atomic
  density, the phase dynamics will still have to be considered as a
  dynamical process. In particular we show that the combination of
  adiabatic and non-adiabatic behaviour can be used to engineer phase
  superposition states.
\end{abstract}


\maketitle

Studying the wave nature of localised single particles allows fundamental questions of quantum mechanics to be addressed. Recently
experimental progress has boosted this area and experiments that can
control single particles have become available in various
systems. These include neutral atoms in optical lattices
\cite{Bloch:08,Khudaverdyan:05,Bakr:09} or microscopic dipole traps
\cite{Yavuz:06,Sortais:06}, electrons in quantum dots \cite{Chan:04}
and several others. One of the advantages of ultra-cold atomic systems
is their purity and low-noise environment, which makes them well
suited for applications in quantum metrology and information
\cite{Briegel:00,Deutsch:00}.

Developing a robust toolbox for engineering using the laws of quantum
mechanics is therefore an important challenge. Compared to classical
physics, many applications require coherent evolution not to be
disturbed and one common process needed is a mechanism that allows
transfer of particles between different trapping sites. Tunneling is
such a mechanism, however in its direct application it leads to Rabi
oscillations which make controlled experiments difficult
\cite{Mompart:03}. Recently a new method, termed coherent tunneling
adiabatic passage (CTAP), has been suggested \cite{Eckert:04,
  Greentree:04, Eckert:06, Rab:08}, which is analogous to the three
level techniques of STIRAP in optical physics \cite{Bergmann:98}. This
technique allows for high fidelity transport between different
trapping sites, with adiabaticity as the only requirement.

While STIRAP is a well investigated technique in optical systems
\cite{Bergmann:98}, its translation into the atom optical realm offers
many new and interesting degrees of freedom to be explored. Recently a
number of studies have focussed on the effects of non-linear dynamics
on the transfer process \cite{Rab:08,Graefe:06}. Here we add another
degree of freedom by studying states with non-trivial phase
distributions and show that the CTAP process is not robust with
respect to conserving the phase and therefore the functional form of
the density distribution. However, we also show that the process can
still be used to control the phase of the quantum state in
question. Phase engineering has become an important technique in the
area of quantum computing recently and the convenience of adiabatic
techniques is that if the associated energy eigenvalue is zero, one
can make use of the usually much smaller geometrical phases
\cite{Moller:07, Moller:08, Unanyan:99}. While this is true for
optical systems, we will show that one has to be more careful in
atom-optical settings.

Tunnelling of an individual vortex is an interesting problem in a
number of systems, including Bose-Einstein condensates and Josephson
junctions. The escape of a single vortex from a pinning potential in a
Josephson junction has been investigated experimentally
\cite{Wallraff:03} and recently the tunnelling of a vortex in
Bose-Einstein condensate has been the subject of numerical study for
double well potentials \cite{Salgueiro:09}. Salgueiro {\sl et al.}
found that the topological defect is preserved on tunnelling of the
condensate and, in fact, can be replicated in such a way that each
potential minimum has a single vortex \cite{Salgueiro:09}.

In the following we will first give a brief introduction to the CTAP
idea and show that the standard approximation of a three level system
is good for the density transport. We will then compare this
approximation to the full integration of the Schr\"odinger equation
for the problem and identify the problems for phase stability. Finally
we will extend the work to look at systems with small non-linearities
and conclude.

\section{Coherent Transport}

\begin{figure}[tb]
  \includegraphics[width=\linewidth,clip=true]{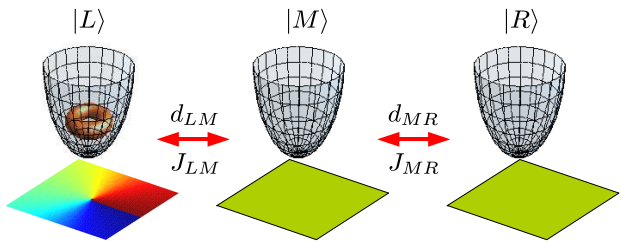}
  \caption{(Color online) Schematic setup for CTAP for an atomic
    state carrying one quantum of angular momentum. The atom is
    initially located in the trap on the left hand side and all other
    traps are considered empty. Tunneling only occurs between nearest
    neighbour traps.}
\label{fig:Schematic}
\end{figure}

The CTAP process for cold atoms considers a system of three
micro-traps, between which a single particle can tunnel (see
Fig.~\ref{fig:Schematic}).  The strength of the tunneling is
determined by the distance and barrier height between the individual
traps. If one assumes all traps to be of the same shape, i.e.~have
resonant energy levels, and only allows for adiabatic processes, one
can write the Hamiltonian in terms of the asymptotic eigenstates of
the individual traps, $|L\rangle,|M\rangle$ and $|R\rangle$, as
\begin{align}
  \label{eq:Hamiltonian}
  H=
  \begin{pmatrix}
    0 & J_{LM} & 0\\
    J_{LM} & 0 & J_{MR}\\
    0 & J_{MR} & 0
  \end{pmatrix}\;.
\end{align} 
The $J_{ij}$ are the tunneling matrix elements between neighbouring
traps and we assume the tunneling between non-neighbouring traps to be
negligible. The eigenstates and eigenvalues of this Hamiltonian are
well known \cite{Bergmann:98} and we will focus here on the so-called
dark eigenstate given by
\begin{align}
  |D\rangle=\cos\theta|L\rangle+\sin\theta|R\rangle\;.
\end{align}
This state only has an indirect contribution from the central
trap through the mixing angle, $\tan\theta=J_{MR}/J_{LM}$, and its
energy eigenvalue is given by $E_D=0$. If one allows for
time-dependent tunneling rates, this state can be used to transfer a
particle from, say, trapping site $|L\rangle$ to $|R\rangle$ with
large fidelity, even in the presence of noise \cite{Eckert:04,
  Greentree:04, Eckert:06}. Note that during this transfer the
particle has no probability to ever be in the state $|M\rangle$.

\begin{figure}[tb]
  \includegraphics[width=\linewidth,clip=true]{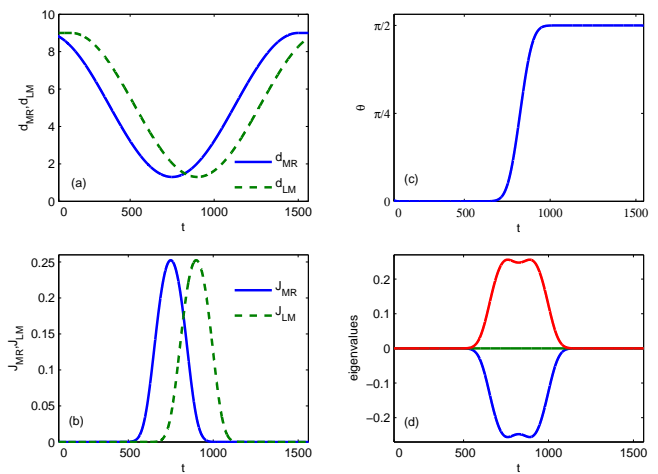}
  \caption{(Color online) (a) distance between the left-middle and
    the middle-right trap. (b) tunneling strength between neighbouring
    traps derived from the analytical model described in the text. (c)
    mixing angle $\theta$ and (d) time-dependent eigenvalues of the
    Hamiltonian \eqref{eq:Hamiltonian}}
\label{fig:3Level}
\end{figure}

To remind the reader, let us briefly review this transfer process: the
tunneling frequencies can become functions of time either through a
time-dependent variation of the distance between the individual traps,
through a modulation of the respective barrier heights or through a
combination of both of them. Here we will assume that the trap
positions change in time and, since we will assume piecewise harmonic
potentials, this will also lead to a decrease in barrier height. If we
therefore first decrease the distance $d_{MR}$ and, with a delay, the
distance $d_{LM}$, we create the familiar counter-intuitive STIRAP
timing sequence for the values of the tunneling strengths $J_{MR}$ and
$J_{LM}$ (see Fig.~\ref{fig:3Level} (a) and (b) and calculations
below). During this process the mixing angle $\theta$ changes from $0$
to $\pi/2$ (see Fig.~\ref{fig:3Level} (c)), which allows a particle
initially trapped on the left hand side to be transferred to the trap
on the right hand side. This is the essence of the celebrated STIRAP
technique.

To calculate the tunneling frequency between two traps as a function
of trap distance, let us turn to the exactly solvable model system of
piecewise harmonic traps \cite{Razavy:03,TF}, which also guarantees
good approximate resonance between the levels at any point in time
\begin{align}
  V=
  \begin{cases}
    \frac{1}{2m}\omega^2\left(x+\frac{d}{2}\right)^2&
    \quad\text{for}\quad x\le 0\\
    \frac{1}{2m}\omega^2\left(x-\frac{d}{2}\right)^2&
    \quad\text{for}\quad x\ge 0\;.
  \end{cases}
\end{align}
The eigenfunctions of this potential are known to be given by
parabolic cylinder functions \cite{Abramowitz}
\begin{align}
  \psi_1(x)&=N_1D_\nu\left[-\frac{2m\omega}{\hbar}(x+\frac{d}{2})\right]&&
   \quad\text{for}\quad x\le 0\;,\\
  \psi_2(x)&=N_2D_\nu\left[ \frac{2m\omega}{\hbar}(x-\frac{d}{2})\right]&&
   \quad\text{for}\quad x\ge 0\;,
\end{align}
where $N_1$ and $N_2$ are the normalisation constants. The
eigenenergies are given by $E/\hbar\omega=\nu+\frac{1}{2}$ and the
quantum numbers, $\nu$, are determined by requiring that the
logarithmic derivatives of the two wavefunctions are equal at $x=0$
\begin{equation}
  \left(\frac{\psi_1'}{\psi_1}\right)_{x=0}=
  \left(\frac{\psi_2'}{\psi_2}\right)_{x=0}\;.
\end{equation}
By solving this condition we can exactly calculate the tunneling
strength between the two traps at any time during the adiabatic
process. If we then diagonalise eq.~\eqref{eq:Hamiltonian} we find the
eigenvalues displayed in Fig.~\ref{fig:3Level}(d), where the dark
state with the eigenvalue of zero is clearly visible. Since we assume
that the whole process is carried out adiabatically, the system will
be in this state at any point in time.

\subsection{Angular Momentum State}

In order to write down the Hamiltonian \eqref{eq:Hamiltonian} for a
realistic system a number of approximations must hold.  First, the
eigenstates of the isolated traps have to be in close resonance, and
second, the levels of the individual traps have to be chosen such that they are well
isolated from all other available states in the system. The second
condition can almost always be fulfiled by making the process more
adiabatic and the first one translates into the simple requirement
that all traps have the same trapping frequency (in case of harmonic
traps). For realistic traps, however, this is problematic, as
potential forces often add when they start to overlap. Several
solutions have been proposed for this, including the use of
time-dependent compensation potentials \cite{Eckert:06}. Here we
assume that this is experimentally possible, as it allows us to
isolate the physics relevant to the dynamics of the phase.

To examine the influence of the CTAP dynamics on the phase
distribution of a quantum state let us first carry out a full
numerical integration of the Schr\"odinger equation for a general
system. This will at the same time function as a control mechanism for
the approximations made above, in particular the fact that we
neglected non-nearest neighbour coupling and treated tunneling as a
one-dimensional process. For numerical simplicity, however, we will
only consider a two-dimensional setup here, as this will allow us to
capture the main physical processes. In scaled co-ordinates
Schr\"odinger's equation is therefore given by
\begin{align}
 i\frac{d}{dt}\psi(x,y) =\left(-\frac{1}{2}\nabla^2
                               +\frac{1}{2}V(x,y)\right) \psi(x,y)\;,
\end{align}
where $V(x,y)$ is the trapping potential of the three traps in the
linear configuration and which fulfils the conditions outlined
above. For generality, all lengths are scaled with respect to the
ground state size of the individual harmonic traps,
$a_0=\sqrt{\hbar/m\omega}$, where $m$ is the mass of the particle and
$\omega$ the trapping frequency of the individual harmonic oscillator
potentials.  All energies are scaled in units of the harmonic
oscillator energy, $E_0=\hbar\omega$.

\begin{figure}[tb]
  \includegraphics[width=\linewidth,clip=true]{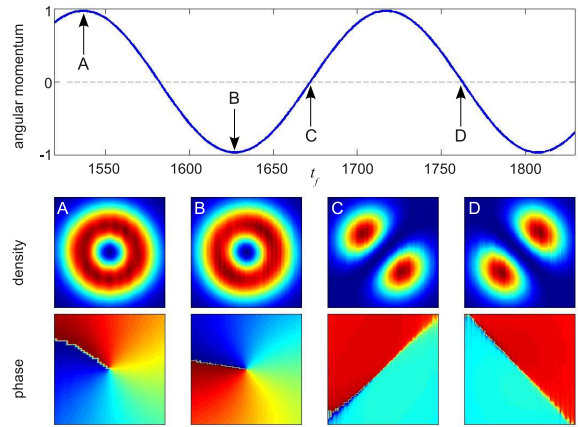}
  \caption{(Color online) Angular momentum as a function of the
    overall duration of the process (upper plot) and the final states
    in the rightmost trap at points A, B, C, and D (lower plots).}
\label{fig:AngularMomentum}
\end{figure}

To be specific, let us choose an initial state that carries a single
quantum of angular momentum and therefore has a phase distribution
that increases by $2\pi$ for a closed loop around the centre of the
state. This state is initially located in the trap on the left hand
side. The results of the numerical integration are summarised in
Fig.~\ref{fig:AngularMomentum} and show surprising dynamics: while
the process still leads to a 100\% transfer of the probability
amplitude to the trap on the right hand side (not shown), the angular
momentum of the final state oscillates continuously between clockwise
and counter-clockwise depending on the overall duration of the process
(see upper part of the figure). Four examples of this change in phase
and density associated with four different durations are shown in the
lower half of the figure. In the first example (A) the system is in
exactly the same state as it started out with, whereas in (B) the
circulation of the flow has been reversed by the CTAP process. The
plots (C) and (D) show the situation where the final state is in a
superposition between clockwise and counter-clockwise rotation, which
also leads to a strongly modified density distribution.

This behaviour might seem surprising at first, as angular momentum is
usually thought of as a conserved quantity. However, in
non-rotationally symmetric geometries this conservation law does not
hold and in fact can lead to interesting dynamics for vortices in
anisotropic potentials \cite{GarciaRipoll:01,Damski:03,Watanabe:07,
  PerezGarcia:07}. As in our example the particle is tunneling between
rotationally symmetric potentials, it is not a priori clear where the
necessary asymmetry comes from.

To gain insight into this behaviour, let us return to the three state
model and consider the transport of a particle in a harmonic
oscillator potential carrying one unit of angular momentum. In two
dimensions its energy is given by $E_\circ=2\hbar\omega$ and the state
is doubly degenerate, $\psi_{n_xn_y}=\psi_{10}$ and
$\psi_{n_xn_y}=\psi_{01}$. We therefore have to write the most general
wavefunction as the superposition
\begin{align}
  \label{eq:AngularMomentumState}
  \psi(x,y)&=\varphi_1(x)\varphi_0(y)+i\varphi_0(x)\varphi_1(y)e^{-i\theta}\;,
\end{align} 
where the one-dimensional, single particle eigenfunctions of the
harmonic oscillator for the ground and first excited state are given
by $\varphi_n$. We have also allowed for a relative phase $\theta$
between the two degenerate states, however we can set this initially
to zero without loss of generality.

As our traps are arranged in a linear configuration along the $y$-axis
we can assume that the dynamics in the different spatial directions
decouple. Tunneling therefore needs to be taken into account only for
the wavefunction part in the $y$-direction and we find that the
Hamiltonian can be split into one for the ground state parts of the
wavefunction and one for the excited states
\begin{align}
  H_0=
  \begin{pmatrix}
    \epsilon_0      & J_0^{LM} & 0\\
    J_0^{LM} & \epsilon_0      & J_0^{MR}\\
    0          & J_0^{MR} &\epsilon_0
  \end{pmatrix}\;,
\end{align}
and
\begin{align}
  H_1=
  \begin{pmatrix}
    \epsilon_1       & J_1^{LM} & 0\\
    J_1^{LM} & \epsilon_1       & J_1^{MR}\\
    0           & J_1^{MR} &\epsilon_1
  \end{pmatrix}\;.
\end{align}
Unlike in the previous section, we cannot simply set the diagonal
elements equal to zero as now both the ground state and first excited energies are involved. However, each of the Hamiltonians has still a dark
eigenstate with the eigenvalues $\epsilon_0$ and $\epsilon_1$,
respectively. If initially a single particle is in the trap on the
left hand side in the state given by
eq.~\eqref{eq:AngularMomentumState}, after the CTAP process, the
Hamiltonians above lead to
\begin{align}
  \varphi_0^L(x)&\longrightarrow\varphi_0^R(x)e^{-i\frac{1}{2}\omega t_f}\\
  \varphi_1^L(x)&\longrightarrow\varphi_1^R(x)e^{-i\frac{3}{2}\omega t_f}\\
  \varphi_0^L(y)&\longrightarrow\varphi_0^R(y)e^{-i\int_0^{t_f}\epsilon_0(t')dt'}\\
  \varphi_1^L(y)&\longrightarrow\varphi_1^R(y)e^{-i\int_0^{t_f}\epsilon_1(t')dt'}\;,
\end{align}
where $t_f$ is the overall duration of the process and the
superscripts $L$ (left) and $R$ (right) refer to the traps the
wavefunction is localised in. Note that although we have used scaled energy units, we reintroduce $\omega$ into the  above expressions for clarity.  One can immediately see that if
$\epsilon_0$ and $\epsilon_1$ are not independent of time, the
wavefunction acquires a relative phase between the two degenerate
states that is dependent on the overall duration of the CTAP process
\begin{align}
  \psi(x,y;t_f)&=
     \varphi_1^R(x)\varphi_0^R(y)
     e^{-i\left[\frac{3}{2}\omega t_f+\int_0^{t_f}\epsilon_0(t')dt'\right]}\nonumber\\
     &+i\varphi_0^R(x)\varphi_1^R(y)
     e^{-i\left[\frac{1}{2}\omega t_f+\int_0^{t_f}\epsilon_1(t')dt'\right]}\;.
\end{align}
For easier understanding let us rewrite the above state by defining a
global and a relative phase as
\begin{align}
\label{eq:gamma}
  \gamma&=\frac{3}{2}\omega t_f+\int_0^{t_f}\epsilon_0(t')dt'
\end{align}
\begin{align}
\label{eq:theta}
  \theta&=-\omega t_f+\int_0^{t_f} [\epsilon_0(t')-\epsilon_1(t')]dt'\;,
\end{align}
so that we can write the final state as
\begin{align}
  \psi(x,y;t_f)=e^{-i\gamma}\left[
     \varphi_1^R(x)\varphi_0^R(y)
    +i\varphi_0^R(x)\varphi_1^R(y)e^{i\theta}\right]\;.
\end{align}
One can clearly see from this that any slight deviation in the
difference between the asymptotic energy levels of the individual
traps will have a significant effect on the final wavefunction. As, in
particular, in realistic situations the shape of the individual
potentials most likely strongly changes, one can expect to be unable
to control the final form of the wavefunction. The CTAP process is
therefore unstable with respect to states with non-trivial phase
distributions \cite{2D}. However, as our example shows, this does not
have to be a random process and in fact it can be used to engineer the
phase of the wavefunction deterministically: by slightly changing the
overall time of the process, one can cycle through all possible states
for the fixed energy $E_\circ=2\hbar\omega$.

\begin{figure}[tb]
  \includegraphics[width=\linewidth,clip=true]{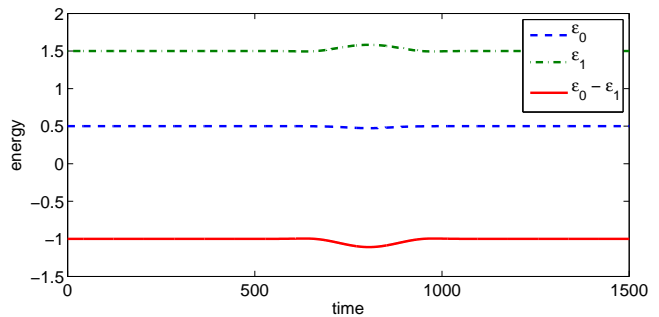}
  \caption{(Color online) Evolution of $\epsilon_0$ (blue dashed), $\epsilon_1$ (green dot-dashed) and $\epsilon_0-\epsilon_1$ (red solid) over the course of the CTAP process. Both energies are initially separated by $\omega=1$, but do not maintain this separation due to the modification of the potential as the traps move closer together.}
\label{fig:en}
\end{figure}

Since in our simulations the potentials are piecewise harmonic, it is
not immediately clear where the change in energies comes from. Let us
therefore in the following carefully examine our model and determine
the influence of various other degrees of freedom. The crucial point
is that during the CTAP process the energy eigenstates in the
$y$-direction slightly change, since the potentials we are considering
are only piecewise harmonic. By bringing them closer together, the
height of the barrier between them changes, leading to a different
asymptotic eigenstate. In fact, the eigenstates are very close to the
parabolic cylinder functions defined above. As the first excited
eigenstate is closer to the local maximum separating the traps, it
will be shifted differently compared to the ground state. While the
difference might only be small, the integration over a long, adiabatic
time interval will lead to a large value for the integral. To demonstrate this, we show in Fig.~\ref{fig:en} the energies for an atom in the state $\varphi_0(x)$ and $\varphi_1(x)$ during the CTAP process. It can be clearly seen that at the time of approach the energy eigenvalues slightly changes, as assumed above. In addition, we show that their difference, which is the integrand of Eq.~\ref{eq:theta}, is not constant and thus gives rise to the time dependant $\theta(t_f)$.

\begin{figure}[tb]
  \includegraphics[width=\linewidth,clip=true]{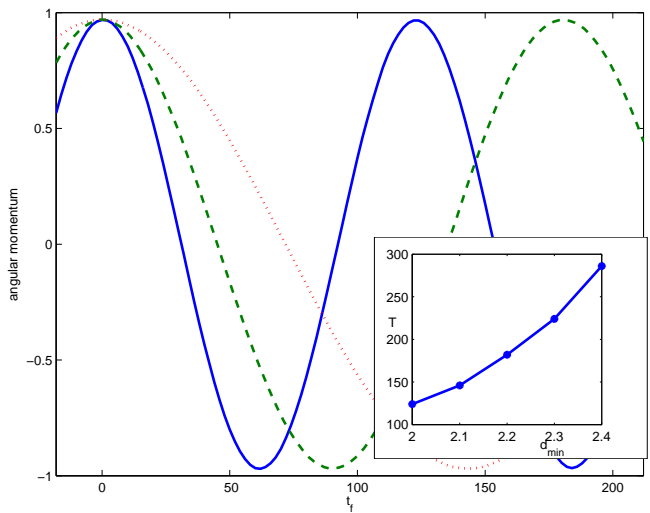}
  \caption{(Color online) Final angular momentum as a function of time
    for different minimum distances between the traps
    $d_\text{min}=2.0$ (full line), $d_\text{min}=2.2$ (dashed line)
    and $d_\text{min}=2.4$ (dotted line). The offset of the different
    curves has been shifted for clarity. The inset shows the increase
    of the period with increasing $d_\text{min}$.}
\label{fig:DMIN}
\end{figure}

From the above argument it follows that the effect should not only
depend on the overall time of the process, but also on the minimum
distance to which the traps approach each other. In
Fig.~\ref{fig:DMIN}, we show the resulting oscillations of angular
momentum for a number of different values for $d_{min}$. As the
minimum distance is increased, the period of the oscillations
increases as well (see inset of Fig.~\ref{fig:DMIN}), which is in
accordance with the explanation given above: a larger minimum distance
between the traps means that the energy levels are deviating less from
the asymptotic values for perfect harmonic potentials.

\subsection{Non-linear CTAP}
\label{sec:NLCTAP}

Let us finally discuss the effect of a non-linearity on the evolution
of the phase. The paradigm of an atomic non-linear system of well
defined phase is a Bose-Einstein condensate and its dynamics can be
described by the so-called Gross-Pitaevskii equation
\begin{align}
   H\psi(x,y)=\left(-\frac{1}{2}\nabla^2+V(x,y)+\frac{U}{2}|\psi|^2\right) 
              \psi(x,y)\;.
\end{align}
Here $U$ is a measure for the non-linearity, which is related to the
scattering strength between the atoms \cite{Pethick:08}.  While the
CTAP process requires resonances of the three asymptotic ground state
at any time, non-linear samples break this due to the time dependence
of a chemical potential, $\mu$, during the tunneling process. For
large chemical potentials the whole process therefore defaults and one
has to resort to different techniques for restoring the resonance
during the process \cite{Rab:08,Nesterenko:08}. As we are only
interested in the phase dynamics, we will restrict ourselves to only
small non-linearities ($\mu\ll\hbar\omega$), for which about 100\%
transfer can still be reached.

\begin{figure}[tb]
  \includegraphics[width=\linewidth,clip=true]{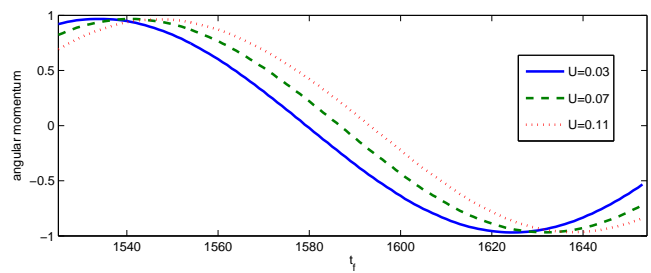}
  \caption{(Color online) Angular momentum as a function of the
    overall time the CTAP process takes for different
    non-linearities. }
\label{fig:NL}
\end{figure}

In Fig.~\ref{fig:NL} we show the effect a small, but increasing
non-linearity has on the final phase. The first thing to notice is
that the oscillatory behaviour of the angular momentum does not get
immediately suppressed by the non-linearity, and that the only effects
are an offset and a small reduction in periodicity (not shown). We
therefore speculate that the effect described in this work can also be
used to create vortex superposition states in Bose-Einstein
condensates in a controlled way, as long as the samples are only
weakly interacting. As it is well known that vortex oscillations in an
anisotropic potential can be suppressed for large enough
non-linearities \cite{GarciaRipoll:01,Watanabe:07}, it might be
interesting to study this effect in the presence of resonance
restoring techniques.

\subsection{Conclusion}

We have shown that the spatial CTAP process for atoms in microtraps
does not conserve the wavefunction form for states with non-trivial
phase distribution. This is due to an unavoidable small time
dependence of the asymptotic eigenstates of the individual traps,
which results from the overlap between them. As any currently
suggested realistic systems shows large deformations due to the
overlap, our work directly applies to current experimental setups.

At the same time we have shown that this instability behaves
deterministically with respect to a change in the overall time of the
CTAP process and can therefore be used to create well defined angular
momentum superposition states. Such states can have applications in
quantum information and have recently seen a surge in interest~\cite{Kapale:05,Thanvanthri:08}. We
have also shown that the presented process even survives in the
presence of small non-linearities, allowing therefore to superpose
vortices in Bose-Einstein condensates.

\begin{acknowledgements}
  The work was supported in part by Science Foundation Ireland under
  project number 05/IN/I852 and by Perimeter Institute for Theoretical
  Physics (SC). Research at Perimeter Institute is supported by the
  Government of Canada through Industry Canada and by the Province of
  Ontario through the Ministry of Research \& Innovation. JB thanks the
  SFI STAR programme for support.
\end{acknowledgements}



\begin{thebibliography}{99}

\bibitem{Bloch:08} I.~Bloch, J.~Dalibard, and W.~Zwerger,
  Rev.~Mod.~Phys.~\textbf{80}, 885 (2008).

\bibitem{Khudaverdyan:05} M.~Khudaverdyan, W.~Alt, I.~Dotsenko,
  L.~F\"orster, S.~Kuhr, D.~Meschede, Y.~Miroshnychenko, D.~Schrader,
  and A.~Rauschenbeutel, Phys.~Rev.~A {\bf 71}, 031404(R) (2005).

\bibitem{Bakr:09} W.S.~Bakr, J.I.~Gillen, A.~Peng, S.~F\"olling and
  M.~Greiner, Nature {\bf 462}, 74 (2009).

\bibitem{Yavuz:06}D.D.~Yavuz, P.B.~Kulatunga, E.~Urban, T.A.~Johnson,
  N.~Proite, T.~Henage, T.G.~Walker, and M.~Saffman,
  Phys.~Rev.~Lett.~{\bf 96}, 063001 (2006).

\bibitem{Sortais:06} Y.R.P.~Sortais, H.~Marion, C.~Tuchendler,
  A.M.~Lance, M.~Lamare, P.~Fournet, C.~Armellin, R.~Mercier,
  G.~Messin, A.~Browaeys, and P.~Grangier, Phys.~Rev.~A {\bf 75},
  013406 (2007).

\bibitem{Chan:04} I.H.~Chan, P.~Fallahi, A.~Vidan, R.M.~Westervelt,
  M.~Hanson, and A.C.~Gossard, Nanotechnology {\bf 15}, 609 (2004).

\bibitem{Briegel:00} H.-J.~Briegel, T.~Calarco, D.~Jaksch, J.I.~Cirac,
  and P.~Zoller, J.~Mod.~Opt.~{\bf 47}, 415 (2000).

\bibitem{Deutsch:00} I.H.~Deutsch, G.K.~Brennen, and P.S.~Jessen,
  Fortsch.~der Phys.~{\bf 48}, 925 (2000).

\bibitem{Mompart:03} J.~Mompart, K.~Eckert, W.~Ertmer, G.~Birkl, and
  M.~Lewenstein, Phys.~Rev.~Lett.~{\bf 90}, 147901 (2003).

\bibitem{Eckert:04} K.~Eckert, M.~Lewenstein, R.~Corbal\'an, G.~Birkl,
  W.~Ertmer, and J.~Mompart, Phys.~Rev.~A {\bf 70}, 023606 (2004). 

\bibitem{Greentree:04} A.D.~Greentree, J.H.~Cole, A.R.~Hamilton, and
  L.C.L.~Hollenberg, Phys.~Rev.~B {\bf 70}, 235317 (2004).

\bibitem{Eckert:06} K.~Eckert, J.~Mompart, R.~Corbal\'an,
  M.~Lewenstein and G.~Birkl, Opt.~Comm.~{\bf 264}, 264 (2006).

\bibitem{Rab:08} M.~Rab, J.H.~Cole, N.G.~Parker, A.D.~Greentree,
  L.C.L.~Hollenberg, and A.M.~Martin, Phys.~Rev.~A {\bf 77}, 061602
  (2008).

\bibitem{Bergmann:98} K.~Bergmann, H.~Theuer, and B.W.~Shore,
  Rev.~Mod.~Phys.~{\bf 70}, 1003 (1998).

\bibitem{Graefe:06} E.M.~Graefe, H.J.~Korsch, and D.~Witthaut,
  Phys.~Rev.~A {\bf 73}, 013617 (2006).

\bibitem{Moller:07} D.~M\o ller, L.~Bojer Madsen and K.~M\o lmer,
  Phys.~Rev.~A {\bf 75}, 062302 (2007).

\bibitem{Moller:08} D.~M\o ller, L.~Bojer Madsen and K.~M\o lmer,
  Phys.~Rev.~A {\bf 77}, 022306 (2008).

\bibitem{Unanyan:99} R.G.~Unanyan, B.W.~Shore and K.~Bergmann,
  Phys.~Rev.~A {\bf 59}, 2910 (1999).

\bibitem{Wallraff:03} A.~Wallraff, A.~Lukashenko, J.~Lisenfeld,
  A.~Kemp, M.V.~Fistul, Y.~Koval, and A.V.~Ustinov, Nature {\bf 425},
  155 (2003).

\bibitem{Salgueiro:09} J.R.~Salgueiro, M.~Zacar\'es, H.~Michinel, and
  A.~Ferrando, Phys.~Rev.~A {\bf 79}, 033625 (2009).

\bibitem{Razavy:03} M.~Razavy, {\sl Quantum theory of tunneling},
  World Scientific (2003).

\bibitem{TF} Note that an approximate formula for the tunneling
  strength between two harmonic traps of the same frequency, $\omega$,
  at a distance $d$ was given in \cite{Eckert:04} as
\begin{align}
  \label{eq:TME}
  \frac{J(d)}{\omega}=\frac{e^{d^2}[1+d(1-\text{erf}(d))]-1}
                           {\sqrt\pi(e^{2d^2}-1)/2d}\nonumber
\end{align}

\bibitem{Abramowitz} M.~Abramowitz and I.A.~Stegun, {\sl Handbook of
    Mathematical Functions with Formulas, Graphs, and Mathematical
    Tables}, Dover (1964).

\bibitem{GarciaRipoll:01} J.J.~Garc\'ia-Ripoll, G. Molina-Terriza,
  V.M.~P\'erez-Garc\'ia, and L.~Torner, Phys.~Rev.~Lett.~{\bf 87},
  140403 (2001).

\bibitem{Damski:03} B.~Damski and K.~Sacha,
  J.~Phys.~A:~Math.~Gen.~{\bf 36}, 2339 (2003).

\bibitem{Watanabe:07} G.~Watanabe and C.J.~Pethick, Phys.~Rev.~A {\bf
    76}, 021605 (2007).

\bibitem{PerezGarcia:07} V.M.~P\'erez-Garc\'ia, M.A.~Garc\'ia-March, and
  A.~Ferrando, Phys.~Rev.~A {\bf 75}, 033618 (2007)

\bibitem{2D} Note that in one dimension this process would only lead
  to a global phase and therefore states with non-trivial phase
  distributions (e.g.~dark solitons) would not be effected.
        
\bibitem{Pethick:08} C.J.~Pethick and H.~Smith {\sl Bose-Einstein
    Condensation in Dilute Gases}, Cambridge University Press; 2
  edition (2008).

\bibitem{Nesterenko:08} V.O.~Nesterenko, A.N.~Novikov, F.F.~de Souza
  Cruz, E.L.~Lapolli, arXiv:0809:5012 (2008).
  
\bibitem{Kapale:05}K.T.~Kapale and J.P.~Dowling, Phys.~Rev.~Lett.~{\bf 95}, 173601 (2005)

\bibitem{Thanvanthri:08} S.~Thanvanthri, K.T.~Kapale, J.~P.~Dowling, Phys.~Rev.~A {\bf 77}, 053825 (2008)

\end{thebibliography}
\end{document}